\documentclass[final]{IEEEtran}
\usepackage{cite}
\usepackage{graphicx}
\graphicspath{{Figs/}}
\usepackage{epsf}
\usepackage{array}
\usepackage{makecell}
\usepackage{amsmath}
\usepackage{amssymb}
\newcolumntype{M}[1]{>{\centering\arraybackslash}m{#1}}
\usepackage{lineno}
\usepackage{fixmath}
\usepackage{hyperref}

\def\BibTeX{{\rm B\kern-.05em{\sc i\kern-.025em b}\kern-.08em
		T\kern-.1667em\lower.7ex\hbox{E}\kern-.125emX}}

\begin{document}
\title{Voice Activity Detection for Transient Noisy Environment Based on Diffusion Nets}
\author{Amir~Ivry,
        Baruch~Berdugo,
        and~Israel~Cohen,~\IEEEmembership{Fellow,~IEEE}% <-this % stops a space
\thanks{This work was supported by the Israel Science Foundation (grant no. 576/16) and the ISF-NSFC joint research program (grant No. 2514/17).}% <-this % stops a space
\thanks{The authors are with the Andrew and Erna Viterbi Faculty of Electrical Engineering,  Technion-Israel Institute of Technology, Haifa 3200003, Israel (e-mail: sivry@campus.technion.ac.il; bberdugo@phnxaudio.com; icohen@ee.technion.ac.il).}% <-this % stops a space
}

\markboth{IEEE JOURNAL OF SELECTED TOPICS IN SIGNAL PROCESSING}%
{Shell \MakeLowercase{\textit{et al.}}: Bare Demo of IEEEtran.cls for IEEE Journals}

\maketitle

\begin{abstract}\label{abstract}We address voice activity detection in acoustic environments of transients and stationary noises, which often occur in real life scenarios. We exploit unique spatial patterns of speech and non-speech audio frames by independently learning their underlying geometric structure. This process is done through a deep encoder-decoder based neural network architecture. This structure involves an encoder that maps spectral features with temporal information to their low-dimensional representations, which are generated by applying the diffusion maps method. The encoder feeds a decoder that maps the embedded data back into the high-dimensional space. A deep neural network, which is trained to separate speech from non-speech frames, is obtained by concatenating the decoder to the encoder, resembling the known Diffusion nets architecture. Experimental results show enhanced performance compared to competing voice activity detection methods. The improvement is achieved in both accuracy, robustness and generalization ability. Our model performs in a real-time manner and can be integrated into audio-based communication systems. We also present a batch algorithm which obtains an even higher accuracy for off-line applications.
\end{abstract}
\begin{IEEEkeywords}
Deep learning, diffusion maps, voice activity detection.
\end{IEEEkeywords}
\section{Introduction}\label{Intro}\IEEEPARstart{V}{oice} activity detection refers to a family of methods that perform segmentation of an audio signal into parts that contain speech and silent parts. In this study, audio signals are captured by a single microphone and contain clean sequences of speech and silence. These signals are mixed with stationary and non-stationary noises (transients), e.g., door knocks and keyboard tapping \cite{DovTransScatter,DovKernel}. Our objective is to correctly assign each captured audio frame into the category of speech presence or absence. A solution to this problem may benefit many speech-based applications such as speech and speaker recognition, speech enhancement, emotion recognition and dominant speaker identification.

In acoustic environments that contain neither stationary or non-stationary noise, speech is detected by using methods that rely on frequency and energy values in short time frames \cite{KrubsackAutocorr,Junqua,VanGerven}. These methods show significant deterioration in performance when noise is present, even with mild levels of signal-to-noise ratios (SNRs). To cope with this problem, several approaches assume statistical models of the noisy signal in order to estimate its parameters \cite{Cho,Chang,Sohn,Ramirez,COH_SP1,COH_SAP1}. Nonetheless, these methods are incapable of properly modeling transient interferences, which constitute an essential part of this study. Ideas that involve dimensionality reduction through kernel-based methods are introduced in \cite{DovDiffusionMaps}, where both supervised and unsupervised approaches have been exploited. However, its main limitation is a significant low-dimensional overlap between speech and non-speech representations. \\ \indent
Machine learning techniques have been employed for voice activity detection in recent studies \cite{Shin,Wu}. In contrast to classic methods, these approaches learn to implicitly model data without assuming an explicit model of a noisy signal. In particular, deep learning based methods have gained popularity in recent years due to a substantial increase in both computational power and data resources. Mendelev et al. \cite{Mendelv} constructed a deep neural network for voice activity detection, and suggested to employ the dropout technique \cite{Srivastava} for enhanced robustness. The main drawback of this method is that temporal information between adjacent audio frames is ignored, due to independent classification of each time frame. Studies presented in \cite{Leglaive,Graves,Hughes,Hong} used a recurrent neural network (RNN) to integrate temporal context with the use of past frames. However, the rapid time variations and prominent energy values of non-stationary noises in comparison to speech are still the main cause of degraded performance in these methods. A recent study conducted by Ariav et al. \cite{Ariav} proposed to use an auto-encoder to implicitly learn an audio signal embedded representation. To enhance temporal relations between frames, this auto-encoder feeds an RNN. Despite its leading performance, the reported results are still unsatisfactory. Our study found that the main limitation of this algorithm is the dense low-dimensional representation forced by the auto-encoder and into the RNN. This density occurs largely due to the joint training of speech and non-speech frames, which fails to enhance their unique features. Thus, their low-dimensional representations, which are the sole information that feeds the RNN, are embedded closely in terms of Euclidean distance. Eventually, this poses a difficulty in separation of speech from non-speech frames based merely on temporal information, which is the core advantage of using RNN architecture.

In this work, we propose an algorithm that addresses the limitations found in the methods proposed in \cite{DovDiffusionMaps} and \cite{Ariav}. We independently learn the low-dimensional spatial patterns of speech and non-speech audio frames through the Diffusion Maps (DM) method. DM is a method that performs non-linear dimensionality reduction by mapping high-dimensional data points to a manifold, embedded in a low-dimensional space \cite{diffMapsIntuition}. The mapped coordinates that lay on this manifold are referred to as DM coordinates. Since this method preserves locality, frames with similar contents in the original high dimension are mapped closely in the low, embedded dimension, with respect to their Euclidean distance. We separately apply DM for speech and non-speech frames through a pair of independent deep encoder-decoder structures. Inspired by the Diffusion nets architecture \cite{DiffusionNets}, the end of each encoder is forced to coincide with the embedded DM coordinates of its high-dimensional input. This approach allows us to differ the intrinsic structure of speech from the ones of transients and background noises based on the Euclidean metric.

We suggest two variations for the voice activity detection algorithm, one for real-time applications and one for batch processes. We test both approaches on five comparative experiments conducted in \cite{DovDiffusionMaps,Ariav,TamuraRobust}. Results show enhanced voice activity detection performance, that surpasses the known state-of-the-art speech detection results. Furthermore, our proposed architecture is more robust and has better generalization ability than competing methods, as demonstrated through experiments.

The remainder of this paper is organized as follows. In Section \ref{ProblemFormulation}, we formulate the problem. In Section \ref{S3}, we introduce the proposed solution. In Section \ref{S4}, we expand on the data set and feature extraction. In Section \ref{S5} we describe the training and testing processes. In Section \ref{S6-Results}, we present the results of the proposed approach for voice activity detection with comparisons to competing methods. Finally, in Section \ref{S7-Conclusion}, we draw conclusions as well as future research directions.
\section{Problem Formulation}\label{ProblemFormulation}
Let ${s}\left[n\right]$ denote the following audio signal:
\begin{align}
{s}\left[n\right] = {s}^{\text{sp}}\left[n\right] + {s}^{\text{st}}\left[n\right] + {s}^{\text{t}}\left[n\right] \textit{,}
\label{eq:fullSignalAddition}
\end{align}
\noindent where $\text{sp, st and t}$ stand for speech, stationary background noise and transient interference, respectively. The time domain signal is processed in overlapping time frames of length $M$. Let $\mathbf{f}_{n}\in{\mathbb{R}^{M}}$ denote the $n\text{th}$ audio frame and let $\{\mathbf{f}_{n}\}_{n=1}^{N}$ denote the audio data set of $N$ time frames. Let $\mathcal{\mathcal{H}}^{0}$ and $\mathcal{\mathcal{H}}^{1}$ be two hypotheses that stand for speech absence and presence, respectively. In addition, let $\mathbb{I}{\left(\mathbf{f}_{n}\right)}$ be a speech indicator of the $n\text{th}$ audio frame, defined as:
\begin{align}
\mathbb{I}{\left(\mathbf{f}_{n}\right)} = \left.
\begin{cases}
1, & \mathbf{f}_n\in\mathcal{H}^{1}\\
0, & \mathbf{f}_n\in\mathcal{H}^{0}\\
\end{cases}
\right\} \textit{.}
\label{eq:hypoDef}
\end{align}
\noindent The goal of this study is to estimate $\mathbb{I}{\left(\mathbf{f}_{n}\right)}$, i.e., to correctly classify each audio frame $\mathbf{f}_{n}$ as a speech or non-speech frame.
\section{Proposed Algorithm for Voice Activity Detection}\label{S3}
Our proposed approach comprises several steps, as illustrated in Fig. \ref{fig: architecture}. Initially, feature extraction is employed in the time domain. The features include the Mel Frequency Cepstral Coefficients (MFCCs) and their low-dimensional representation, generated by the DM method. A detailed description is given in Section \ref{S4.2}. Subsequently, a deep encoder-decoder based neural network is used to learn the unique patterns of speech and non-speech signals. Since this structure makes use of the DM method, it is regarded in this study as diffusion encoder-decoder (DED). Next, error measures are extracted from the deep architecture. Those errors are represented in a coordinate system, notated in this study as error map. It should be highlighted that no mathematical operation is applied on the errors extracted from the network. i.e., the error map is merely a representation form which allows us to conduct better analysis and gain deeper insights on the performance of our detector, as will be shown along this paper. A classifier, fed by the coordinates of the error map, is constructed to separate speech presence and absence. In this study, two different modes are used for classification. First, a batch mode is considered, where a substantial corpus of speech and non-speech audio frames must be at hand, in order to evaluate the outcome of the DM process correctly. In batch mode, both low and high-dimensional errors are taken into account during classification. The second classification mode is real-time, which exploits merely high-dimensional error information. In this case, integration of DM is not required, which allows a frame-by-frame classification with negligible delay.
\subsection{Deep Encoder-Decoder Neural Network}\label{S3.1}
Our approach suggests that speech frames can be separated from non-speech frames based on their intrinsic low-dimensional representation. Ideas from \cite{DiffusionNets} are adopted to merge DM with two independent, identically constructed DEDs, notated by $\text{DED}^{i}$, where $i\in\{0,1\}$. DM allows a geometric interpretation of the data by constructing its underlying embedding, which can be represented by the middle layer of any basic encoder-decoder network \cite{Ariav}. To exploit this property, the middle layer is forced to coincide with the true DM coordinates of the input layer. As a result, the encoder of $\text{DED}^{i}$ is trained to map spectral features affiliated with $\mathcal{H}^{i}$ from their original space to the lower diffusion space. Subsequently, the decoder of $\text{DED}^{i}$ learns the inverse mapping back to the high-dimensional feature space.

\begin{figure}[!t]
	\centering\includegraphics[width=3.8in]{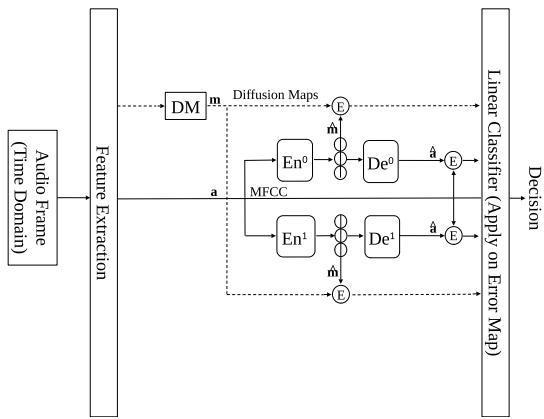}
	\caption{Proposed architecture for voice activity detection. Dashed line is valid only for batch mode and solid line is constantly employed for both batch and real-time modes. `En' and `De' are abbreviations for encoder and decoder, respectively. Superscripts 0 and 1 relate to the index of the corresponding trained DED. The circled `E' notation refers to an error calculation unit, defined in (\ref{eq:defineTwoAnomalies}).}
	\label{fig: architecture}
\end{figure}

A deep architecture is constructed to implement the above notion, as illustrated in Fig. \ref{fig: architecture}. In this proposed system, each DED comprises two stacked parts, an encoder and a decoder. The former is constructed from a 72 neurons input layer followed by two layers of 200 neurons each and a final layer of 3 neurons. The deep decoder is a reflection of the deep encoder. While the size of the middle and hidden layers are determined empirically, the size of the input (and thus, the output) layer of each DED is derived from the feature extraction process, as described in Section \ref{S4.2}. In the output of each layer, an identical activation function is employed on each neuron (\ref{eq:actFun}).
\subsection{Error Maps and Voice Activity Detection Classifier}\label{S3.2}
Let us denote a single observation of an input feature vector as $\textbf{a}$ and its true DM coordinates as $\textbf{m}$. Additionally, $\hat{\textbf{m}}$ and $\hat{\textbf{a}}$ denote the encoding of $\textbf{a}$ by a trained encoder and its reconstruction by a trained decoder, respectively. Each observation is fed into the trained DEDs simultaneously. That way, the relations between each hypothesis and the constructed embeddings are compared under the same conditions. These measures are employed through $e_{\mathrm{en}}\left({\textbf{m}}\right)$ and $e_{\mathrm{de}}\left({\textbf{a}}\right)$, where:
\begin{align}
e_{\mathrm{en}}\left({\textbf{m}}\right) = \Vert{\textbf{m}-\hat{\textbf{m}}}\Vert_{1} \textit{ ; } e_{\mathrm{de}}\left({\textbf{a}}\right) = \Vert{\textbf{a}-\hat{\textbf{a}}}\Vert_{1} \textit{,}
\label{eq:defineTwoAnomalies}
\end{align}
\noindent while $\Vert{\cdot}\Vert_{1}$ denotes the $\ell_{1}$ norm. Namely, as $e_{\mathrm{en}}\left({\textbf{m}}\right)$ represents the mapping error, $e_{\mathrm{de}}\left({\textbf{a}}\right)$ is associated with the reconstruction error of $\textbf{a}$.

In this study, two classification modes are considered. In the batch mode, both $e_{\mathrm{en}}\left({\textbf{m}}\right)$ and $e_{\mathrm{de}}\left({\textbf{a}}\right)$ are taken into account. Namely, each observation $\textbf{a}$ ultimately generates two pairs of errors, one from each DED. These errors are interpreted as a four-dimensional coordinate that is embedded into an error map. In the real-time mode, on the other hand, only the decoder error $e_{\mathrm{de}}\left({\textbf{a}}\right)$ is extracted from each DED. i.e., in this scenario a two-dimensional coordinate is embedded into the error map.

Subsequently, a support vector machine (SVM) classifier with linear kernel is trained on the error map, which contains the generated error measures from a corpus of observations. The objective of this classifier is to separate between coordinates affiliated with different hypotheses. As a result, two decision regions are created, for speech presence and absence. Since $\text{DED}^{i}$ is trained to construct a low-dimensional manifold on which $\mathcal{H}^{i}$ is embedded, frames related to $\mathcal{H}^{i}$ highly fit the learned mapping of $\text{DED}^{i}$. This leads to substantially lower errors, which could be easily identified as a separate cluster. This assumption is derived from the property of the DM method, in which the diffusion distance in the original feature dimension is proportional to the $\ell_{1}$ norm in the diffusion space. In this study, a classic SVM classifier is shown to be sufficient.

It is worth noting that we have also implemented an alternative architecture to the one presented in Fig. \ref{fig: architecture}, which involves a unified network instead of an SVM. The goal of this was to assert the improvement, and thus justify the employment, of our suggested system over the alternative of the fully connected neural network, which is commonly used in deep learning algorithms. We concatenated the output layer of both $\text{DED}$ branches to each other and to the input layer. Then, this augmented layer was connected to a single-bit output neuron that carries the VAD decision. Results have shown very similar performance, with a slight tendency to the SVM based method. As a results, we have decided to use the originally presented architecture. Two minor advantages of the SVM can be noted over the unified neural network. First, it is less computationally expensive in comparison to using an additional hidden layer, which will consume higher memory and time during back propagation. Second, the original method explicitly constructs the error measures and feeds them to the SVM, which leads to high separation of speech from silence. Therefore, the hidden layer attempts to implicitly represent the data in a similar manner, i.e., to find the relation between the neurons which will ultimately lead to good separation. Representing the error measures in the two-dimensional space and applying the SVM on it both serves as a more natural, intuitive classification algorithm and avoids the infamous ``black box'' property of the neural network, as well as grants us the ability to analyze the decision of the detector in a helpful and profound manner, as will be done later on.
\section{Database and Feature Extraction}\label{S4}
\subsection{Database}\label{S4.1}
We adopt the audio database presented in \cite{DovDiffusionMaps} to construct a DED training set, a classifier training set and a back to end test set. This database is obtained from 11 different speakers reading aloud an article chosen from the web, while making natural pauses every few sentences. Naturally, these recordings are composed of sequences of speech followed by non-speech frames. Each sequence varies from several hundred milliseconds to several seconds in length. These signals were recorded with an estimated SNR of 25~dB at a sampling rate of 8 $\text{kHz}$. Each of the 11 signals is 120 seconds long and it is processed using short time frames of 634 samples with 50\% overlap, which effectively generates a 25 frames/second rate. The clean speech signal $s^{\text{sp}}\left[n\right]$, defined in (\ref{eq:fullSignalAddition}), is used to determine the presence or absence of speech in each time frame, and to construct a label set accordingly.

These clean audio signals are contaminated by 42 different pairs of additive stationary and non-stationary noises, which construct a varied data set. The noise signals employed include white and colored Gaussian noise, babble and musical instruments. Transients include keyboard taps, scissors snapping, hammering and door knocks.
\subsection{Feature Extraction}\label{S4.2}
We wish to exploit the ability of deep neural networks to learn complex relations between their inputs and outputs. Hence, our objective is to feed our architecture with features that express the unique patterns of each hypothesis $\left(\ref{eq:hypoDef}\right)$. To generate spectral information from the time domain database, MFCCs are employed. These coefficients are concatenated along a fixed number of adjacent frames, in order to gain temporal context between them. DM is applied to integrate spatial properties and to find a relation between the spectrum of the signal and its geometric low-dimensional structure.
\subsubsection{Mel Frequency Cepstral Coefficients}\label{S4.2.1}
Features based on a spectral representation of audio signals are fully adopted from the study of Dov et al. \cite{DovDiffusionMaps}. To construct them, weighted MFCCs are employed. MFCCs use the perceptually meaningful Mel-frequency scale, which allows a compact representation of the spectrum of speech \cite{BLogan}. MFCC features are used in the presence of highly non-stationary noise, where they were found to perform well for speech detection tasks \cite{Mousazadeh}.

However, speech frames may have similar MFCC representation to frames comprising highly non-stationary noise as well, since they both have akin spectral attributes. To address this challenge, noise estimation is performed in each frame and the MFCCs in that frame are weighted accordingly \cite{DovDiffusionMaps,Mousazadeh,DavisMerm}. This enables better analysis by separating the background noise from the rest. Next, several consecutive time frames are taken into account. Hence, the nature of transients, which their typical duration is assumed to be of the order of a single time frame, can be exploited.

Formally, consider $\textbf{a}_{n}\in\mathbb{R}^{C}$ as a row vector of $C$ coefficients, consisting of weighted MFCCs, and their first and second derivatives, $\Delta$ and $\Delta\Delta$, respectively. These values are extracted from the $n$th time domain audio frame $\mathbf{f}_{n}$, introduced in Section \ref{ProblemFormulation}. Let:
\begin{align}
\underline{\mathbf{a}}_{n} = \left[\textbf{a}_{n-J},\dots,\textbf{a}_{n},\dots,\textbf{a}_{n+J}\right]\in\mathbb{R}^{\left(2J + 1\right)C} \textit{}
\end{align}

\noindent denote concatenation of feature vectors from $2J+1$ adjacent frames, where $J$ is the number of past and future time frames. For $J\geq1$, the elements of $\underline{\mathbf{a}}_{n}$ in the presence of transients are expected to vary faster than in the presence of speech.

In this study, the number of MFCCs is $8$, as commonly used. Thus, $\textbf{a}_{n}$ comprises of $C= 24$ coefficients. For practical considerations, we assign a relatively small value of $J=1$. This allows informative characterization of audio frames based on past-future relations, while consuming low computational load. Thus:
\begin{align}
\underline{\mathbf{a}}_{n} = \left[\textbf{a}_{n-1},\textbf{a}_{n},\textbf{a}_{n+1}\right]\in\mathbb{R}^{72} \textit{.}
\label{eq:tempFeatWFrames}
\end{align}

\noindent Next, standardization is applied on (\ref{eq:tempFeatWFrames}). Let us assume a set of $N$ observations, while the $n$th observation is given by (\ref{eq:tempFeatWFrames}), for $n\in\{1,\ldots,N\}$. For each feature index $l\in\{1,\ldots,72\}$, a row vector $\textbf{O}_{l}\in\mathbb{R}^{N}$ is defined as:
\begin{align}
\textbf{O}_{l} = \left[\underline{\mathbf{a}}_{1}\left(l\right),\dots, \underline{\mathbf{a}}_{N}\left(l\right)\right] {.}
\label{eq:defineObservation}
\end{align}
Then, the mean and standard deviation of $\textbf{O}_{l}$ are extracted and termed ${\mu}_{l}$ and ${\sigma}_{l}$, respectively. Next, the following vectors are constructed:
\begin{align}
\boldsymbol{\mu}=\left[{\mu}_{1},\dots,{\mu}_{72}\right]\text{ ; } \boldsymbol{\sigma}=\left[{\sigma}_{1},\dots,{\sigma}_{72}\right]\textit{.}
\label{eq:defineStandardParams}
\end{align}
Let $\underline{\tilde{\mathbf{a}}}_{n}(l)$ denote the $l$th element of the standardized $\underline{\mathbf{a}}_{n}(l)$, defined as:
\begin{align}
\underline{\tilde{\mathbf{a}}}_{n}(l)=\frac{\underline{\mathbf{a}}_{n}(l) - \mu_{l}}{\sigma_{l}}
\textit{.}
\label{eq:defineStandardProcess}
\end{align}
\subsubsection{Diffusion Maps}\label{S4.2.2}
The middle layer of any basic autoencoder architecture can be viewed as a low-dimensional representation of its input layer \cite{AutoEncoders}. Our method exploits this by forcing the middle layer to coincide with the embedded coordinates of $\underline{\tilde{\mathbf{a}}}_{n}$, generated by the DM method \cite{CoifmanLafon}. Thus, the encoder learns to approximate this low-dimensional mapping, while the decoder learns the inverse high-dimensional mapping. DM is a manifold learning approach that is established on the graph Laplacian of the high-dimensional data corpus \cite{coifmanLafonHarmonics}. DM has been employed well in several signal processing, image processing and machine learning applications \cite{LafonKeller,Farbman,Singer,TalmonGannot,DavidDataOrg,Gapshtein,Mishne,Haddad,CoifmanHirn}.

Let us consider a set of feature vectors $\{\underline{\tilde{\mathbf{a}}}_{n}\}_{n}$, constructed according to (\ref{eq:defineStandardProcess}). A weighted graph is created with the elements of the set as nodes (or points), where the weight of the edge connecting these nodes is given by the commonly used radial basis function (RBF) kernel. The scaling parameter of the kernel is set separately for each edge as in \cite{Zelnik}. Practically, merely the 10 nearest neighbors of every point are used to compute the edges. Namely, edges that are not among the nearest neighbors of $\underline{\tilde{\mathbf{a}}}_{n}$ are nullified.

In order for the embedding and the distribution of the nodes to be independent, we perform normalization of the data. Therefore, an approximation of the Laplace-Beltrami operator on the data is obtained \cite{CoifmanLafon,LafonKeller}. This operation generates a row-stochastic matrix $P$ which can be viewed as the transition matrix of a Markov chain on the data set $\{\underline{\tilde{\mathbf{a}}}_{n}\}$. Two sets of bi-orthogonal left and right eigenvectors, $\{\mathbold{\phi}_{n}\}$ and $\{\mathbold{\psi}_{n}\}$, are constructed by employing an Eigenvalue decomposition of $P$. This process also yields a series of strictly positive eigenvalues $1=\vert\lambda_{0}\vert\geq\vert\lambda_{1}\vert\geq...\geq\vert\lambda_{n-1}\vert>0$.

Through informal experiments, we found that for retaining the desired patterns of speech and non-speech frames, it is sufficient that the embedded dimension is set to $d=3$ (excluding the trivial dimension associated with $\lambda_{0}$). Furthermore, $d$ is small enough to exclude undesired high frequency noise, mostly represented by higher dimensions. The low-dimensional embedding of $\underline{\tilde{\mathbf{a}}}_{n}$ (\ref{eq:defineStandardProcess}) is notated by $\mathbf{m}_{n}$ and defined as:
\begin{align}
\mathbf{m}_{n} = \left(\lambda_{1}\mathbold{\psi}_{1}\left(\underline{\tilde{\mathbf{a}}}_{n}\right),\ldots,\lambda_{3}\mathbold{\psi}_{3}\left(\underline{\tilde{\mathbf{a}}}_{n}\right)\right) \textit{.}
\end{align}
\noindent Therefore, the set $\{\underline{\tilde{\mathbf{a}}}_{n}\}$ is embedded into the Euclidean space $\mathbb{R}^{3}$. In this space, the Euclidean distance is equal to the diffusion distance in the high-dimensional space of $\{\underline{\tilde{\mathbf{a}}}_{n}\}$.

Our architecture integrates an activation function which maps its input to the interval $\left[0,1\right]$. On the other hand, $\mathbf{m}_{n}$ often holds values which may exceed this interval. Therefore, this mismatch increases the error measures defined in (\ref{eq:defineTwoAnomalies}). Earlier works have demonstrated that prediction accuracy can be improved by normalizing DM coordinates \cite{Bridle}. We employ these notions to overcome the aforementioned mismatch, by mapping the dynamic range of $\mathbf{m}_{n}$ to $\left[0,1\right]$. Specifically, the transformation that is employed corresponds to connecting $\mathbf{m}_{n}$ to $\tilde{\mathbf{m}}_{n}$ through a softmax layer \cite{softmaxIntro}, as follows:
\begin{align}
\tilde{\mathbf{m}}_{n}\left(k\right)=\frac{e^{\mathbf{m}_{n}\left(k\right)}}{\sum_{l=1}^{3}e^{\mathbf{m}_{n}\left(l\right)}} \textit{,}
\label{eq:MappingOnDM}
\end{align}
\noindent where $1\leq k\leq 3$. As a result, $0\leq\tilde{\mathbf{m}}_{n}\left(k\right)\leq1$ and $\sum_{k=1}^{3}\tilde{\mathbf{m}}_{n}\left(k\right)=1$.

\section{Experimental Setting}\label{S5}
\subsection{Notation}\label{S5.0}
Let $\textbf{s}_{j}\in\mathbb{R}^{L}$ denote the contaminated audio signal associated with speaker $j\in\{1,\ldots,11\}$, comprising of $L$ samples. Let ${\textbf{s}^{i}_{j}}$ denote the union of audio time frames in ${\textbf{s}_{j}}$ that belong to hypothesis $\mathcal{H}^{i}$. Then, $\underline{\textbf{s}}^{i}$ is defined as the concatenation of ${\textbf{s}^{i}_{j}}$ with respect to all 11 speakers, namely:
\begin{align}
\underline{\textbf{s}}^{i} = \left[\textbf{s}^{i}_{1},\ldots,\textbf{s}^{i}_{11}\right]\textit{,}
\label{eq:sameHypoAll}
\end{align}
\noindent where $i\in\{0,1\}$.
\subsection{DED Training Process}\label{S5.1}
Let us consider the two distinct sets $\underline{\textbf{s}}^{0}$ and $\underline{\textbf{s}}^{1}$. Two training sets, notated  $\underline{\textbf{s}}^{0}_{\text{tr,ded}}$ and $\underline{\textbf{s}}^{1}_{\text{tr,ded}}$, are created by randomly extracting 70\% of $\underline{\textbf{s}}^{0}$ and $\underline{\textbf{s}}^{1}$, respectively. Following Section \ref{S4.2.1}, the feature vector extracted from the $n$th frame of $\underline{\textbf{s}}^{i}_{\text{tr,ded}}$ is denoted $\underline{\mathbf{a}}^{i}_{\text{tr,ded},n}\in~\mathbb{R}^{72}$. Next, standardization process (\ref{eq:defineStandardProcess}) is applied on the latter, which yields $\underline{\tilde{\mathbf{a}}}^{i}_{\text{tr,ded},n}\in~\mathbb{R}^{72}$. This reveals two advantages; first, the network performs a faster learning process. This occurs since standardization implicitly weights all features equally in their representation. Thus, the rate at which the weights connected to the input nodes learn is balanced. This balance allows to rescale the learning rate through the learning process. As a result, the adaptive gradient descent optimization method can be deployed instead of the traditional gradient descent. Second, this approach reduces saturation effects, caused by large values assigned to activation functions. Next, the DM method is applied on the set $\{\tilde{\underline{\mathbf{a}}}^{i}_{\text{tr,ded},n}\}_{n}$ separately, for each $i\in\{0,1\}$, as described in Section \ref{S4.2.2}. The resulting low-dimensional embedding is clipped to the dynamic range $\left[0,1\right]$ and denoted by $\tilde{\mathbf{m}}^{i}_{\text{tr,ded},n}\in~\mathbb{R}^{3}$. The proposed architecture entails that while $\underline{\tilde{\mathbf{a}}}^{i}_{\text{tr,ded,n}}$ is fed to $\text{DED}^{i}$, the middle layer of the latter is enforced to coincide with $\tilde{\mathbf{m}}^{i}_{\text{tr,ded},n}$. Let us denote $\text{DED}^{i}_{\text{tr}}$ as the $i$th trained DED.

We integrate the Positive Saturating Linear Transfer (PSLT) activation function, defined as follows:
\begin{align}
\sigma(z) = \left.
\begin{cases}
0, & z\leq0\\
z, & 0\leq z\leq 1\\
1, & z\geq1\\
\end{cases}
\right\} \textit{.}
\label{eq:actFun}
\end{align}
\noindent The dynamic range that $\sigma(z)$ generates, differently from the known ReLU, suggests maintaining the fluctuations which may appear along the tangled network. Employing $\sigma(z)$ is beneficial in terms of low computational load that is consumed during back propagation, since the derivative of $\sigma(z)$ is simply 1 or 0, neglecting singularities. During back propagation, a nullified derivative will decrease computation time even further, at the expense of updating the weights of the network with less information. Empirically, it was shown not to deteriorate performance. Also, it should be highlighted that complex non-linear patterns can still be learned by the deep architecture. Pre-training is applied on each layer separately in an unsupervised manner, using encoder-decoder structures with 1 epoch and learning rate of 0.1. The optimized weights obtained by this process are considered instead of the random initialization commonly used, which enhances performance since it helps the network to avoid local minima. Pre-training is extremely effective in case there is a relatively small amount of training data, as in our scenario. Next, fine-tuning is applied separately on the encoder and the decoder. Namely, $\underline{\tilde{\mathbf{a}}}^{i}_{\text{tr,ded,n}}$ is encoded into a low-dimensional representation and decoded back to the output layer independently. Subsequently, the two tuned parts are merged and fine-tuning is again utilized, this time on the full stacked DED. Optimization is employed by back propagation through time, which makes use of gradient descent method, parameterized with learning rate of $10^{-5}$ and momentum of 0.9. Prior to pre-training, the weights are initialized with values drawn from a random normal distribution with zero mean and variance 0.01. Cost function with $L_{2}$ weight regularization of $10^{-7}$, sparsity regularization of 4 and sparsity proportion of 0.1 is employed. Relatively large sparsity related parameters were assigned, to achieve two goals. First, this allows the networks to avoid over-fitting by effectively ignoring weights with negligible values. Second, it decreases the computational load, since the embedding process involves a sparse affinity matrix. The network was trained until either 1,000 epochs or minimum gradient value of $10^{-6}$ were achieved. A typical simulation as such took approximately 10 hours on a i7-7820HQ CPU 64-bit operating system, x64 based processor. \\
\indent In this study, the architecture was trained using a batch size of 128 observations. As a result, less memory was used compared with feature-by-feature feeding, since fewer registers were employed at the same time. Moreover, the training was accelerated due to less updates performed, i.e., less propagations through the network. On the other hand, batch training may lead to less accurate and stable estimation of the gradient.
\subsection{Classifier Training Process}\label{S5.2}
Let $\underline{\mathbf{s}}^{i}_{\text{tr,cl}}$ contain random 15\% of the observations contained in $\underline{\mathbf{s}}^{i}$, and let $\underline{\mathbf{s}}_{\text{tr,cl}} = ~ \left[\underline{\mathbf{s}}^{0}_{\text{tr,cl}}, \underline{\mathbf{s}}^{1}_{\text{tr,cl}}\right]$ be the full classifier training set. $\underline{\mathbf{s}}_{\text{tr,cl}}$ is built so it is disjoint with the DED training set. Similarly to the DED training process, $\underline{\tilde{\mathbf{a}}}_{\text{tr,cl},n}\in\mathbb{R}^{72}$ and $\tilde{\mathbf{m}}_{\text{tr,cl},n}\in\mathbb{R}^{3}$ represent feature vectors extracted from the $n$th frame of $\underline{\mathbf{s}}_{\text{tr,cl}}$, according to (\ref{eq:defineStandardProcess}) and (\ref{eq:MappingOnDM}), respectively.

Error measures are defined to distinguish between features that are mapped and reconstructed well and features that are not. Consider two outcomes of propagating $\tilde{\underline{\mathbf{a}}}_{\text{tr,cl},n}$ through $\text{DED}^{i}_{\text{tr}}$. Namely, its low-dimensional predicted representation, denoted by $\tilde{\mathbf{m}}^{i}_{\text{pr},n}$, and its subsequently predicted reconstruction, denoted by $\tilde{\underline{\mathbf{a}}}^{i}_{\text{pr}, n}$. Consequently, the following error measures are defined, given $\underline{\tilde{\mathbf{a}}}_{\text{tr,cl},n}$:
\begin{align}
e_{\text{en}}^{i}(n)\triangleq\Vert\tilde{\mathbf{m}}_{\text{tr,cl},n}-\tilde{\mathbf{m}}^{i}_{\text{pr},n}\Vert_{1}\textit{,}
\label{eq:EncoderError}
\end{align}
\noindent which is associated with $\text{encoder}^{i}$, and:
\begin{align}
e_{\text{de}}^{i}(n)\triangleq\Vert \underline{\tilde{\mathbf{a}}}_{\text{tr,cl},n}-\underline{\tilde{\mathbf{a}}}^{i}_{\text{pr}, n}\Vert_{1}\textit{,}
\label{eq:DecoderError}
\end{align}
\noindent associated with $\text{decoder}^{i}$. In both cases, $\Vert{\cdot}\Vert_{1}$ denotes the $\ell_{1}$ norm.

According to (\ref{eq:EncoderError}) and (\ref{eq:DecoderError}), it can be inferred that a pair of numerical errors is generated by feeding $\underline{\tilde{\mathbf{a}}}_{\text{tr,cl},n}$ to each trained DED. In this study, the two pairs of errors, associated with $\text{DED}_{\text{tr}}^{0}$ and $\text{DED}_{\text{tr}}^{1}$, are interpreted as a coordinate in $\mathbb{R}^{4}$ and are represented by $\left(e_{\text{en}}^{0}(n), \text{ } e_{\text{de}}^{0}(n), e_{\text{en}}^{1}(n), e_{\text{de}}^{1}(n)\right)$. Namely, $\underline{\tilde{\mathbf{a}}}_{\text{tr,cl},n}$ is eventually represented in a four-dimensional coordinate system.

An SVM classifier, notated by $\mathcal{C}$, is applied on the error map, as detailed in Section \ref{S3.2}. In this study, $\mathcal{C}$ is trained to separate coordinates held by $\mathcal{H}^{0}$ from coordinates held by $\mathcal{H}^{1}$ (\ref{eq:hypoDef}). Thus, two decision regions are created. In this study, both real-time and batch modes are considered, as described in Section \ref{S5.3}. For batch mode, $\mathcal{C}$ is trained on both the encoder and decoder errors projected on the error map, i.e., $\mathcal{C}$ is a three-dimensional hyper plane, embedded in $\mathbb{R}^{4}$. Real-time mode only exploits the decoder error. Namely, in this case the error map is a two-dimensional coordinate system, and correspondingly $\mathcal{C}$ divides $\mathbb{R}^{2}$ into two regions.
\subsection{Testing Process}\label{S5.3}
The DM method requires a batch of both speech and non-speech frames to estimate the low-dimensional embedding. This is impractical for real-time mode where a very small number of frames is available. Therefore, two testing processes are presented; a frame-by-frame testing process in which employment of the DM method is not required, and a batch testing process, which is shown to be more accurate, with substantially higher delay.
\subsubsection{Batch Mode Testing Process}\label{S5.3.1}
In batch mode, both the encoder and the decoder errors are exploited, which increases prediction accuracy. On the other hand, the encoder error is well approximated as long as a large batch of time domain audio data from both hypotheses (\ref{eq:hypoDef}) is at hand, which leads to delay in prediction.
The test set, notated by $\underline{\mathbf{s}}_{\text{te}}$, is constructed by following similar steps as in the previous section, while ensuring that the intersection of $\underline{\mathbf{s}}_{\text{te}}$ and the training sets of the DED neural network and classifier is empty. $\underline{\mathbf{s}}_{\text{te}}$ includes 15\% of both $\underline{\mathbf{s}}^{0}$ and $\underline{\mathbf{s}}^{1}$ $\left(\ref{eq:sameHypoAll}\right)$. For completion, $\underline{\tilde{\mathbf{a}}}_{\text{te},n}\in\mathbb{R}^{72}$ and $\tilde{\mathbf{m}}_{\text{te},n}\in\mathbb{R}^{3}$ denote the feature vectors associated with the $n$th observation of $\underline{\mathbf{s}}_{\text{te}}$, extracted according to (\ref{eq:defineStandardProcess}) and (\ref{eq:MappingOnDM}), respectively.

Let $\left(e_{\text{en}}^{i}(n),e_{\text{de}}^{i}(n)\right)$ represent the two-dimensional coordinate generated by the propagation of $\underline{\tilde{\mathbf{a}}}_{\text{te},n}$ through $\text{DED}_{\text{tr}}^{i}$. $e_{\text{en}}^{i}(n)$ and $e_{\text{de}}^{i}(n)$ are produced according to (\ref{eq:EncoderError}) and (\ref{eq:DecoderError}), respectively. For the sake of clarity, we neglect the time index $n$ and address $e_{\text{en}}^{i}(n)$ and $e_{\text{de}}^{i}(n)$ as a two-dimensional coordinate $\left(e_{\text{en}}^{i},e_{\text{de}}^{i}\right)$. As stated earlier, $\left(e_{\text{en}}^{0},e_{\text{de}}^{0}\right)$ and $\left(e_{\text{en}}^{1},e_{\text{de}}^{1}\right)$ are concatenated and projected into a four-dimensional error map. Let $\text{R}_{j}$ stand for region $j$ created by the devision $\mathcal{C}$ applied to the error map, where $j\in\{0,1\}$. Ultimately, the following decision rule is applied by the classifier $\mathcal{C}$ on the input feature vector $\underline{\tilde{\mathbf{a}}}_{\text{te},n}$:
\begin{align}
\mathcal{C}\{\underline{\tilde{\mathbf{a}}}_{\text{te},n}\} = \left.
\begin{cases}
\mathcal{H}^{0}, & \left(e_{\text{en}}^{0},e_{\text{de}}^{0},e_{\text{en}}^{1},e_{\text{de}}^{1}\right)\in\text{R}_{0}\\
\mathcal{H}^{1}, & \left(e_{\text{en}}^{0},e_{\text{de}}^{0},e_{\text{en}}^{1},e_{\text{de}}^{1}\right)\in\text{R}_{1}\\
\end{cases}
\right\} \textit{.}
\label{eq:decisionRuleOffline}
\end{align}

\subsubsection{Real-Time Mode Testing Process}\label{S5.3.2}
Since immediate prediction is often required in many audio-based applications, real-time mode is considered as the main branch of this study. Compared with the batch mode, the low-dimensional error is now unavailable. Meaning, the high-dimensional error becomes the single measure to distinguish between audio frames of different hypotheses.

Let $e_{\text{de}}^{i}(n)$ denote the error produced by propagating $\underline{\tilde{\mathbf{a}}}_{\text{te},n}$ through $\text{DED}_{\text{tr}}^{i}$. In a similar manner to the batch mode, $e_{\text{de}}^{0}(n)$ and $e_{\text{de}}^{1}(n)$ are joined and projected into a two-dimensional error map. For sake of clarity, we again address these two measures as $(e_{\text{de}}^{0},e_{\text{de}}^{1})$. Let $\text{R}_{j}$ stand for region $j$ created by the devision $\mathcal{C}$ applied to the two-dimensional error map, where $j\in\{0,1\}$. As a result, the following decision rule is considered by the classifier $\mathcal{C}$, regarding input feature vector $\underline{\tilde{\mathbf{a}}}_{\text{te},n}$:
\begin{align}
\mathcal{C}\{\underline{\tilde{\mathbf{a}}}_{\text{te},n}\} = \left.
\begin{cases}
\mathcal{H}^{0}, & \left(e_{\text{de}}^{0},e_{\text{de}}^{1}\right)\in\text{R}_{0}\\
\mathcal{H}^{1}, & \left(e_{\text{de}}^{0},e_{\text{de}}^{1}\right)\in\text{R}_{1}\\
\end{cases}
\right\} \textit{.}
\label{eq:decisionRuleOnline}
\end{align}
\begin{figure}[!t]
	\centering\includegraphics[width=3.5in, height=1.83in]{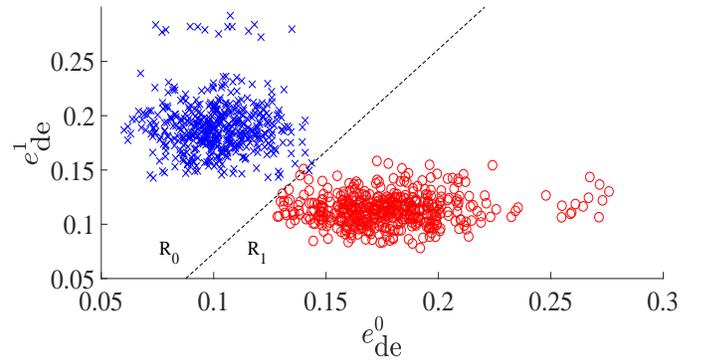}
	\caption{Two-dimensional error map, generated by the real-time mode. Red circles denote speech presence and blue `x' marks denote speech absence. The trained linear SVM classifier is represented by the dashed line and the decision regions it generates are notated by {\normalfont $\text{R}_{0}$} and {\normalfont $\text{R}_{1}$}.}
	\label{fig: onlineFullPerf}
\end{figure}
\begin{table*}[!t]
	\renewcommand{\arraystretch}{1.3}
	\caption{Comparison Between Voice Activity Detection Methods in Terms of Accuracy Rate, with Respect to the TP+TN \protect\linebreak (True Positive + True Negative) Measure. The Results of the Proposed Method are Highlighted.}
	\label{table: evaluation}
	\centering
	\begin{tabular}{|M{2.5cm}|M{1.7cm}|M{1.7cm}|M{1.7cm}|M{1.7cm}|M{1.7cm}| M{0.4cm}|}
		\hline
		& \thead{Babble \\ 10dB SNR \\ Keyboard} & \thead{Musical \\ 10dB SNR \\ Hammering} & \thead{Colored \\ 5dB SNR \\ Hammering} & \thead{Musical \\ 0dB SNR \\ Keyboard} & \thead{Babble \\ 15dB SNR \\ Scissors} & std \\ \hline
		Tamura & 73.6 & 83.8 & 83.9 & 73.8 & 81.2 & 5.2\\ \hline
		Dov - Audio & 87.7 & 89.9 & 87.8 & 86.5 & 90.2 & 1.6 \\ \hline
		Dov - Video & 89.6 & 89.6 & 89.6 & 89.6 & 89.6 & 0\\ \hline
		Dov - AV & 92.9 & 94.5 & 92.8 & 92.9 & 94.6 & 0.9\\ \hline
		Ariav - AV & 95.8 & 95.4 & 95.9 & 95.1 & 97.2 & 0.8 \\ \hline
		Proposed Real Time & \textbf{98.4} & \textbf{98.3} & \textbf{98.3} & \textbf{98.3} & \textbf{98.5} & \textbf{0.1} \\ \hline
		Proposed Batch & \textbf{99.3} & \textbf{99.6} & \textbf{99.3} & \textbf{99.3} & \textbf{99.5} & \textbf{0.1} \\ \hline
	\end{tabular}
\end{table*}
\section{Experimental Results}\label{S6-Results}
In each of the experiments described in this section, comparisons are made between our proposed approach and several competing voice activity detectors. In order to avoid skewness and unfair imbalance, performances were generated by using identical experimental conditions. Specifically, the same test set, acoustic setup and optimization measure, i.e., $\text{TN}+\text{TP}$ (true positive + true negative), are uniformly employed. To allow appropriate assessment of performances, two measures are used: The optimized TP+TN measure, and the relation between TP and TN measures.
\subsection{Performance of Proposed Approach}\label{S6.1}
\subsubsection{Accuracy}\label{S6.1.1}
Primarily, the proposed method is applied using 100\% of the DED training data set in a batch mode, as detailed in Section \ref{S5.3.1}. The accuracy rate is 99.1\%. In this mode, voice activity is detected by using both low and high-dimensional numerical measures. This performance gives rise to the main assumption of this research. Namely, that speech can be distinguished from transients based on their underlying geometric structures.
Real-time voice activity detection is performed according to Section \ref{S5.3.2}. In this mode, the accuracy rate reaches up to 98.1\% when 100\% of the DED training data set is used. Visualization of the error map is given in Fig. \ref{fig: onlineFullPerf}. It should be highlighted that similar visualization is not given for the batch mode, since the corresponding error map lays in $\mathbb{R}^{4}$. These results reflect on the strong relation between low and high-dimensional information. Namely, even though low-dimensional measures are not integrated into the decision rule, the separation in the diffusion space is implicitly expressed through the inverse mapping of the decoder. Therefore, the reconstructed high-dimensional information in the feature space is a sufficient measure to tell apart speech from non-speech frames. By examining the results, high robustness can be concluded. Namely, despite the variety of stationary and non-stationary noises included in the database, the intrinsic structure of speech is still well detected.
\begin{figure}[!t]
	\centering\includegraphics[width=3.5in]{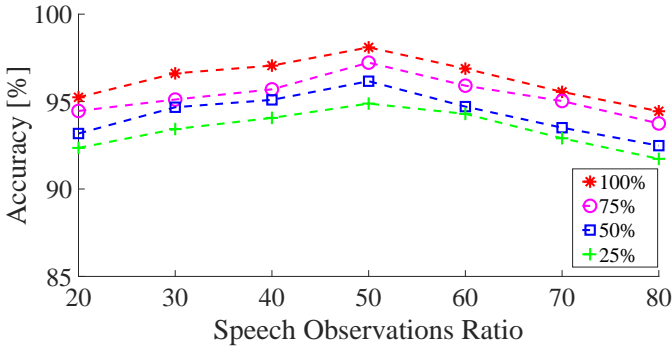}
	\caption{Accuracy rate percentage (TP+TN) of the proposed method using the real-time mode. Different fractions of the full DED training set (25,50,75,100[\%]) are considered along a grid of speech observations ratios.}
	\label{fig: SizeVsRatios}
\end{figure}

\subsubsection{Generalization}\label{S6.1.2}
Generalization and sensitivity of the proposed method are analyzed by performing an additional experiment in the real-time mode. These properties are examined with respect to two parameters; the corpus size of the DED training set and the ratio of speech observations in the latter. In this experiment, 5 different fractions of the full amount of the DED training set are considered. For each fraction, 5 different ratios between speech and non-speech observations are inspected. Results are demonstrated in Fig. \ref{fig: SizeVsRatios}. It can be observed that the accuracy rate surpasses 95\%, even when merely 50\% of the training data is available, which projects on the low sensitivity of the proposed algorithm to this measure. Also, the maximal accuracy is achieved when the speech observations ratio is equal to 50\%, i.e., when there is an equal amount of speech and non-speech observations in the DED training set. This optimal ratio allows the network to learn two separate manifolds with minimal bias. This bias, if exists, can come to surface during testing, when one mapping is more robust than the other. In this case, relying on Euclidean distance between manifolds as done in this research may be harmful for classification. It can also be inferred that the performance has low sensitivity to changes in the speech observations ratio parameter. For example, let us consider the results achieved by exploiting 100\% of the training corpus. Then, speech observations ratios of 20\%, 50\% and 80\% yield accuracies of 95.2\%, 98.1\% and 94.4\%, respectively. It is interesting to note that the degradation in performance is not symmetric around the ratio of 50\%. i.e., degradation is more noticeable when the amount of noise observations is lower than those of speech in the training process. This can be related to the high varying nature of non-stationary noises in comparison to speech. Meaning, larger corpus of transients is needed to construct a robust low-dimensional structure with the DM method.
\linebreak \indent As mentioned in Section \ref{S4.1}, the constructed database comprises 42 different combinations of stationary and non-stationary noises. Thus, a fundamental question is concerned with the ability of the proposed detection system, and specifically $\text{DED}^0$, to generalize well to other types of noise. In order to increase the generalization ability of the suggested detector to noises of various kinds, we performed several actions that regard both the architecture of the system and the feature extraction process. The way the architecture is built puts emphasis on both the difference between speech and noise, and on the similarity of noise to previously trained noises. As a result, the decision mechanism of the system relies on a combination of two learning systems. The features that are extracted from the time domain are constructed to exploit this form of architecture. During training, not only temporal and spectral features are derived, as traditionally done in state-of-the-art methods, but also the informative spatial diffusion map features. This reveals the unique intrinsic geometric structure of speech utterances. Ultimately, when feeding the system with unseen noise, its intrinsic structure is evaluated by the system and compared against speech and non-speech frames separately. Therefore, the performance of the system is not sensitive to unseen noises, in comparison to competing methods, as shown through the experimental setup detailed earlier in this section.

\subsection{Comparison to Competing Methods}\label{S6.2}
\begin{figure}[!t]
	\centering\includegraphics[width=3.5in]{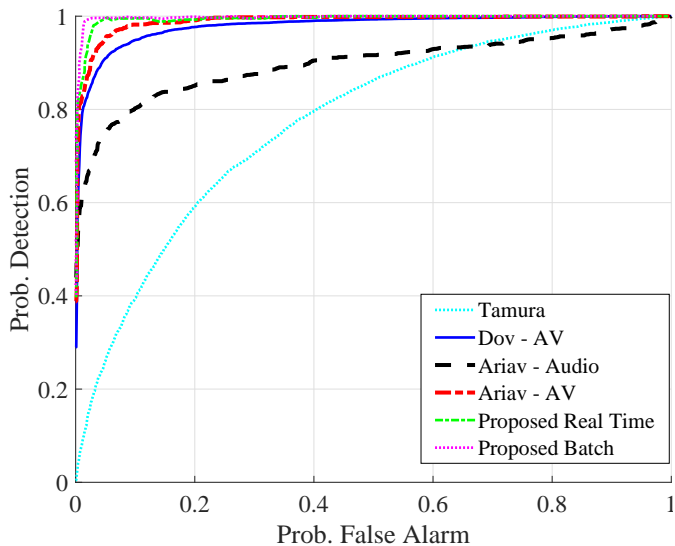}
	\caption{Probability of detection versus probability of false alarm in an acoustic environment of babble noise with 10 dB SNR and keyboard transient interferences.}
	\label{fig: ROC_Babble_10_keyboard5}
\end{figure}

\begin{figure}[!t]
	\centering\includegraphics[width=3.5in]{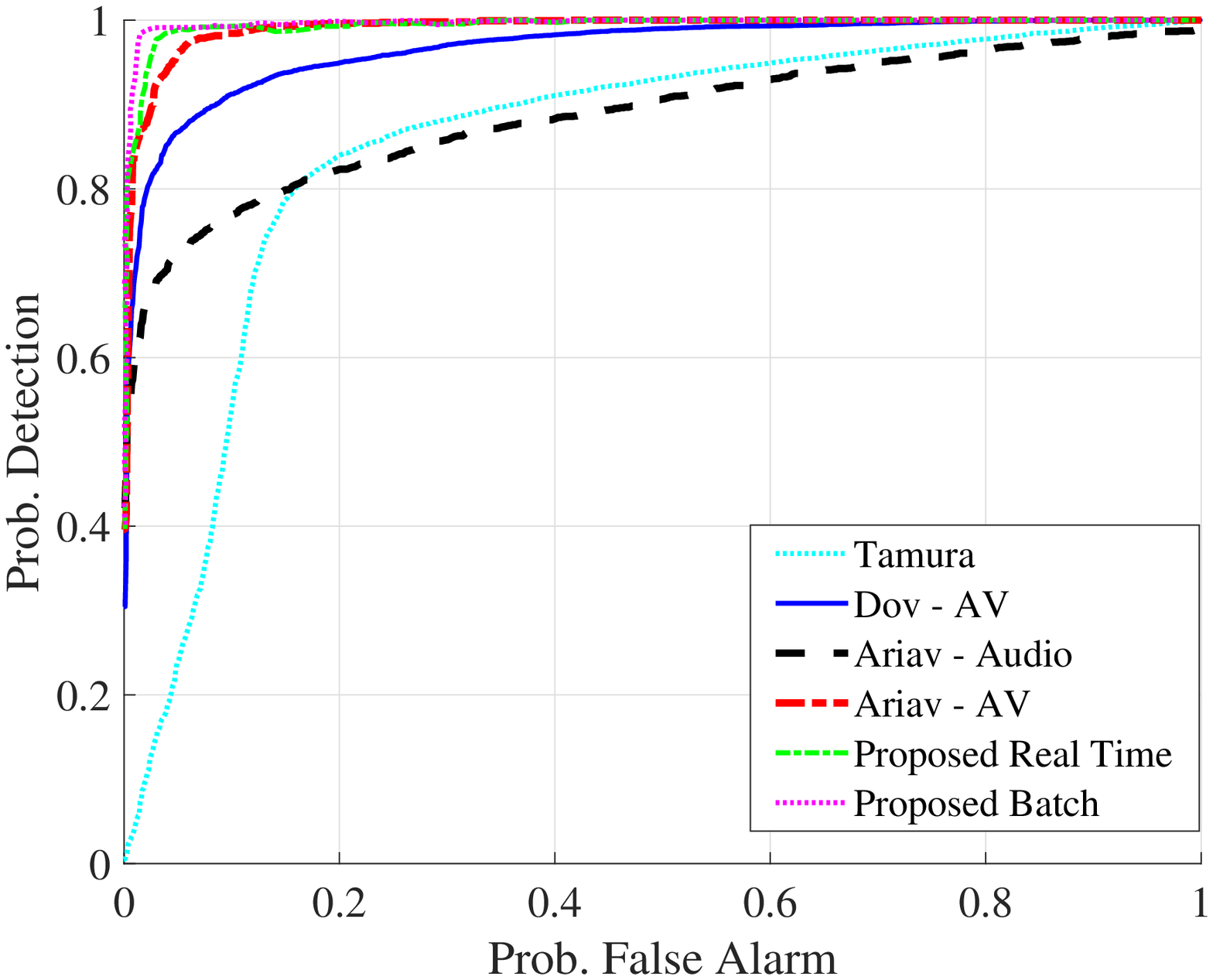}
	\caption{Probability of detection versus probability of false alarm in an acoustic environment of colored noise with 5 dB SNR and hammering transient interferences.}
	\label{fig: ROC_Colored_5_hammering4}
\end{figure}

\begin{figure}[!t]
	\centering\includegraphics[width=3.5in]{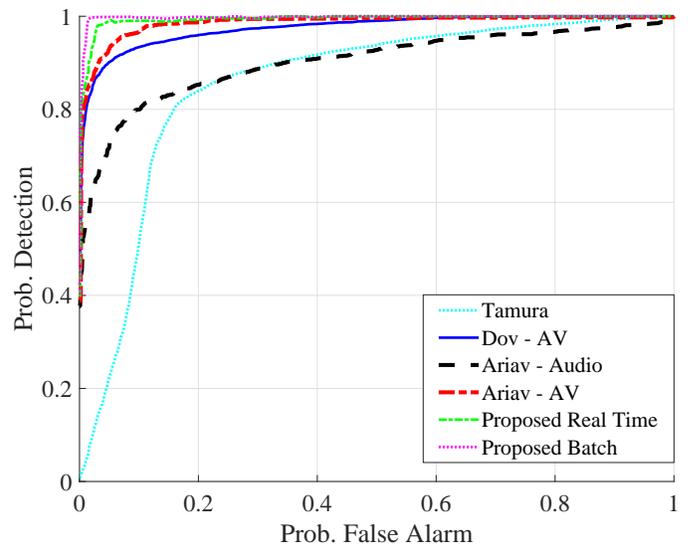}
	\caption{Probability of detection versus probability of false alarm in an acoustic environment of musical noise with 10 dB SNR and hammering transient interferences.}
	\label{fig: ROC_musical10hammering5}
\end{figure}

In order to assert the performance of our architecture in a global scale, it is compared to 5 voice activity detectors. The competing methods are presented in \cite{DovDiffusionMaps,Ariav,TamuraRobust} and are denoted ``Ariav'', ``Dov'' and ``Tamura'', respectively. Table~\ref{table: evaluation} presents the performance of each method in 5 different acoustic environments that compose of transients (keyboard, hammering, scissors) and stationary noises (babble, musical, colored Gaussian noise) with different SNR values (0, 5, 10, 15 [dB]). The real-time and batch modes are notated by `Proposed Real-Time' and `Proposed Batch', respectively. \\ \indent
It can be observed that the proposed algorithm, even in real-time mode, achieves the best accuracy rate through all varied setups. It should be highlighted that the proposed solution exploits only audio signals, while competing methods rely on integration of both audio and video data. \\ \indent
By observing the standard deviation (std) measure in Table~\ref{table: evaluation} it is shown that, unlike competing methods, the performance of the proposed approach is barely affected by the change in the acoustic environments. This high robustness can be related to the construction of intrinsic representations of the audio frames. These representations do not consider the contents of transients or background noises, but merely their intrinsic geometric patterns. These patterns are unique for speech and non-speech audio frames, which allows enhanced performance regardless of the setup. The results presented in Table~\ref{table: evaluation} show slight improvement in comparison to the results presented in Section \ref{S6.1}. While in the former, 5 specific setups are inspected, 37 additional setups are considered in the latter. This indicates the existence of specific combinations of speech, stationary and non-stationary noises that are harder to comprehend. Deeper analysis of this phenomenon should be addressed in future work. \\ \indent
To allow further evaluation, we employ the receiver operating characteristic (ROC) curve. Three acoustic setups presented in Table~\ref{table: evaluation} are considered in Figs. \ref{fig: ROC_Babble_10_keyboard5} -- \ref{fig: ROC_musical10hammering5}. In each ROC curve, the real time and batch proposed approaches are compared against four competing voice activity detectors. Since the test set is identical and balanced across all methods, a constructive comparison is made by the ROC curves. The latter allows analysis of the relation between TP and TN, thus delivering information about the trade-off between the two. It is worth noting that TN can be derived from the false positive (FP) measure, held by the x axis, by simply applying the relation TN = 1-FP. It can be observed that our voice activity detector outperforms the competing methods in a wide range of operating points.

\subsection{Performance Analysis}\label{S6.3}
\begin{figure}[!t]
	\centering\includegraphics[width=3.5in]{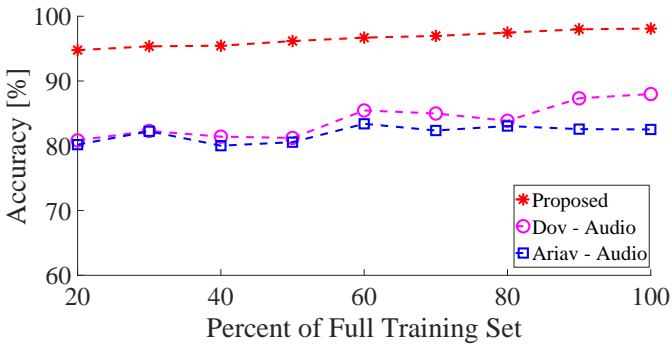}
	\caption{Accuracy rate percentage (TP+TN) of the proposed method using the real-time mode and the methods presented in Dov et al. \cite{DovDiffusionMaps} and Ariav et al. \cite{Ariav}. Performance is presented along a grid of different fractions of the full DED training set, while the speech observations ratio is fixed to 50\%.}
	\label{fig: SizesVsMethods}
\end{figure}

This study presents a voice activity detection method that reaches substantially higher accuracy results in comparison to other state-of-the-art methods. This improvement can be attributed to several novelties, where two of them are considered the most influential. First, the integration of the DM method, forced at the end of the encoder. Second, construction of two separate DEDs, one trained with speech presence observations and the second with speech absence observations. This section is divided into two main parts. Initially, the differences between two competing methods and the proposed approach are analyzed and theoretical explanations of the gap in accuracies are given. Then, two experiments are conducted to establish these explanations. \\ \indent
First, the method proposed in \cite{DovDiffusionMaps} is considered. In this method, low-dimensional embedding is built with the DM method, as done in our study. This embedding is constructed by considering joint relations between speech and non-speech features. However, our approach employs the DM method by considering relations between features of the same hypothesis only. In order to evaluate the influence of this difference on the degradation in performance, the algorithm proposed in \cite{DovDiffusionMaps} has been implemented. Consequently, high overlap of speech and non-speech embeddings in the diffusion space has been observed. This method performs voice activity detection mainly by modeling two low-dimensional Gaussian mixture models. Meaning, this approach aims to separate speech from non-speech coordinates by constructing a separator from a sum of weighted exponential kernels. As a result, overlapped coordinates are highly at risk to be misclassified.\\ \indent
Next, the method proposed in \cite{Ariav} is analyzed. In this approach, a single auto-encoder attempts to learn the low-dimensional embedding of both speech and non-speech frames. As a result, joint embedding is shown to lead to high overlap in the low-dimension, much like in the research conducted in \cite{DovDiffusionMaps}. Additionally, this architecture does not consider the DM method as a constraint on the embedded data, so dimensionality reduction is done automatically. This leads to a lack of spatial information in the low-dimension and absence of geometric insight. Ultimately, this causes significant overlap between low-dimensional representations and to deterioration in performance. The high accuracy shown in \cite{Ariav} can be related to high exploitation of temporal relations, carried by the RNN, and integration of visual features in the classification process. To explore the performance of the network without video, the authors of this work implemented audio-only version of the method presented in \cite{Ariav}. The outcome shows severe degradation in performance, as the average accuracy is 83\% with respect to all 5 setups considered in Table~\ref{table: evaluation}.
\\ \indent Two experiments are conducted in order to validate the above notions.
First, the algorithms proposed in \cite{DovDiffusionMaps} and \cite{Ariav} are implemented with merely audio data, as demonstrated in Fig. \ref{fig: SizesVsMethods}. Accuracy rates of these methods are calculated by employing different fractions of the full DED training set. For this particular experiment, the ratio of speech observations was fixed to 50\%, to achieve optimal results. Several interesting insights can be obtained based on these outcomes. Primarily, there is a substantial gap between performances when considering only the audio data and neglecting visual features. Moreover, it is noticeable that the method proposed in \cite{Ariav} is not affected as much by the change in the amount of training observations. As previously stated, the latter does not consider any geometric or structural constraint on the embedded data. Therefore, as long as the training observations are divided roughly equal between hypotheses, their amount has lower significance. On the other hand, the study presented in \cite{DovDiffusionMaps} highly relies on the intrinsic structure of the data. i.e., the more training observations are available, the better the joint relations between speech and non-speech features are modeled. In this case, larger training set leads to a more robust manifold construction.

\begin{figure}[!t]
	\centering\includegraphics[width=3.5in]{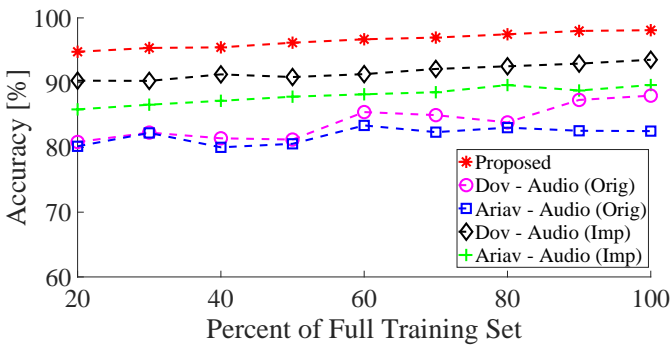}
	\caption{Accuracy rate percentage (TP+TN) of the studies presented in Dov et al. \cite{DovDiffusionMaps} and Ariav et al. \cite{Ariav}, along with the performance of the proposed algorithm in the real-time mode. Accuracy is presented along a grid of different fractions of the full DED training set, while the speech observations ratio is fixed to 50\%. Each of the two competing methods is implemented once in the original form (marked `Orig') and once with integration of concepts from the proposed method (marked `Imp').}
	\label{fig: sizesVsMyImpl}
\end{figure}

\indent In order to further explore the core of the advantages of the proposed approach, another experiment is conducted. This time, the studies in \cite{DovDiffusionMaps} and \cite{Ariav} are implemented by integration of several principles of this study. It should be noted that the detection algorithm presented in each of these studies remains the same. In \cite{DovDiffusionMaps}, the algorithm was altered such that the low-dimensional coordinates are learned separately for speech and non-speech frames before applying the Gaussian mixture model on the generated manifolds.
In \cite{Ariav}, two separate auto-encoders were implemented. Each auto-encoder learned the low-dimensional mapping of speech and non-speech audio frames independently. Also, the DM method was applied in a similar manner to the proposed method in order to integrate spatial information. The output of each encoder was inserted into a separate RNN. The output of each RNN represents the probability that a test observation is taken from a speech audio frame. Ultimately, the probabilities of the two RNNs are intersected and a prediction is made by a constructed decision rule. \\ \indent
The results of this experiment are given in Fig.~\ref{fig: sizesVsMyImpl}. For each method, the accuracy is calculated along a grid of fractions of the full DED training set, while the speech observations ratio is once again set to 50\%. Moreover, the performance of each method is given once with its original implementation and once with the improved implementation that combines principles from our method. Regarding the studies presented in \cite{DovDiffusionMaps,Ariav}, the accuracies of the two new implementations significantly improve. Also, these models are less sensitive to changes in the size of the DED training corpus.

Even though an increase in performance can be observed, the studies presented in \cite{DovDiffusionMaps} and \cite{Ariav} still do not reach the results of the proposed method. The core classification algorithm of each of the three discussed methods remains unchanged through all the comparative experiments conducted in this study. Therefore, the core classification algorithm proposed in our study may be responsible for the observed gap.

\section{Conclusions}\label{S7-Conclusion}
In this work we have performed voice activity detection with audio-based features. We separately represented the low-dimensional geometric structures of speech and non-speech frames by integrating the diffusion maps method with two independent, encoder-decoder based, deep neural networks. This separation of speech from stationary noises and transients during the training process of the two networks also led to high robustness and generalization abilities, as well as low sensitivity to the amount of available training data. The proposed method has shown state-of-the-art results in a real time mode, and can be integrated into dedicated communication systems. Nonetheless, non stationary noises are still the main cause of false detection in this research, due to their high varying nature. This challenge may be addressed by employment of more distinctive geometric features as well as assimilation of joint constraints between the encoder and decoder. It would be instructive to further factorize the proposed approach and analyze the improvement. Moreover, a heuristic explanation regarding the relation between diffusion maps and the presented method can be meaningful for further understanding. One hypothesis, for instance, links between transition in time and on the transition map. Another theory suggests that the corresponding Markov chain is a sequence of phonemes, and the diffusion rate in the diffusion map corresponds to the velocity of phonemes pronunciation. Additionally, the performance of the proposed detection method in reverberant and noisy acoustic environments with signal-to-noise ratios lower than 0 dB, should be explored.

\section*{Acknowledgements}
The authors thank the Guest Editor, Dr. Bo Li, and the anonymous reviewers
for their constructive comments and useful suggestions.

\bibliographystyle{IEEEtran}
\bibliography{IEEEabrv,Refs}

\end{document}